# The Nexus between Job Burnout and Emotional Intelligence on Turnover Intention in Oil and Gas Companies in the UAE

**Anas Abudaqa[a], Mohd Faiz Hilmi[b], Norziani Dahalan[c],** [a,b,c]School of Distance Education, Universiti Sains Malaysia, 11800 Penang, Pulau Pinang, Malaysi[a],

Currently, job satisfaction and turnover intentions are the significant issues for oil and gas companies in the United Arab Emirates (UAE). These issues need to be addressed soon for the performance of the oil and gas companies. Thus, the aim related to the current study is to examine the impact of job burnout, emotional intelligence, and job satisfaction on the turnover intentions of the oil and gas companies in the UAE. The goals of this research also include the examination of mediating the influence of job satisfaction alongside the nexus of job burnout and turnover intentions of the oil and gas companies in the UAE. The questionnaire method was adopted to collect the data from the respondents, and Smart-PLS were employed to analyse the data. The results show that job burnout, emotional intelligence, and job satisfaction have a positive association with turnover intentions. In contrast, job satisfaction positively mediates the nexus between job burnout and turnover intentions. These results provide the guidelines to the policymakers that they should enhance their focus on job satisfaction and turnover intentions of the employees that improve the firm performance.

## Introduction

Satisfied employees work much harder and provide better creative ideas to organisations. This, in turn, helps organisations to develop new ways of earning or new ways of strategically completing work. Job satisfaction is an important aspect that is required to be considered by different firms. However, this is needed to be understood that not only employees but also leaders or managers are crucial parts of any organisation (Chung, Jung, & Sohn, 2017). They need to be satisfied as well so that they can provide better training to employees, and together they can work hard for the success of the organisation. United Arab Emirates (UAE) organisations offer several additional benefits to employees, and, for that reason, it has the highest percentage of satisfied employees. In a recent survey, in oil and gas companies of UAE, nearly 58% of full-time employees rated their work "Good," while almost 12% of employees rated their work "Great" (Iaffaldano & Muchinsky, 1985). This reveals their satisfaction levels at their jobs.

However, this is contrary to other studies that show that employees are satisfied in only a few organisations. Still, there is a large number of organisations that are not able to provide even basic amenities. These organisations fail to understand that benefits and amenities help in gaining not only the best performance from employees but will also help in retaining them for earning revenues and profits (Lawler III & Hackman, 1971). This phenomenon was the primary motivating force to take up the current research study.

In this context, this research study endeavoured to find the relationship between all potential stress factors of organisations including emotional intelligence and satisfaction levels that cause job burnout leading to the turnover intention of business workers. The study took this premise that job burnout is often the result of the failure of employees to adjust and cope up with the work pressure, which is very high in UAE's oil and gas industry (Vanda & Haghighi, 2014). Threats to the job and other severe risks are also present in these oil companies. Due to this, job burnout is seen among employees of the UAE's oil and gas companies.

The reason to include emotional intelligence in this study is justified because it is the sixth most crucial skill any employee requires to stay longer in any working environment (Leunissen, Sedikides, Wildschut, & Cohen, 2018). Employees, as well as managers, must be aware of their skills and capabilities. They must show empathy to others and build strong relationships as well with other staff members. In the UAE, as mentioned above, human beings from different parts of the world come and work in oil and gas companies. Thus, they must have all interpersonal skills, particularly the skills of Emotional Intelligence, to survive in this working environment. Otherwise, they might find themselves in the middle of some bad issues (Ning, Li, Yang, Ge, & Liu, 2014). However, UAE oil companies understand the value of these skills, and therefore only skilled people are recruited.

Leaders or managers of oil and gas companies observe each employee carefully and then only allocate the job to them to achieve the best out of them. Employees have a broad range of nontechnical and technical opportunities that are provided by the company (Safi & Kolahi, 2015). They get training from experienced people, and all of these lead to job satisfaction.

At oil and gas companies in the UAE, employees and managers are recruited after a lengthy recruitment process and a lot is expected from them. Therefore, they possess higher emotional intelligence skills. They have full freedom and opportunities to do what they can, so less job burnout takes place. Despite high satisfaction levels as shared in few studies, employee turnover rate is high in the UAE and is nearly 56% (Griffin, Hogan, Lambert, Tucker-Gail, & Baker, 2010). Employees are not leaving this country, instead, they are switching their jobs due to the low benefits provided to them. A closer look at the job environment of the UAE discloses that not all organisations follow the rules and regulations set by the Government for employees. Thus, due to these few organisations, out of which some are gas and oil companies as well, the employee turnover rate is high in this country.

The UAE is motivated to take care of its employees, but due to immense work pressure, gradually job burnout takes place. There are several fields to work for, but employees prefer to work in oil and gas companies as they pay well. Sometimes earnings are tax-free based on the policy of the home country. However, people from different geographical locations come and work here in all those companies, and this diversity is present.

*Problems Related to Job Burnout*

Often organisations do not value their employees as they think they are paying for getting the job done. However, as stated by Heffernan & Rochford(2017), when employees work in a place where they cannot explore themselves or their ideas, they start feeling bored or suffocated. Mental fatigue overcomes them, which usually turns into job burnout. As a result, employees start showing less focus on work, and the company earnings are impacted. Moreover, often organisations in the UAE force employees to do extra jobs in several fields in which they do not have any experience. Past studies observe that a few oil and gas companies often make teams with employees from different areas to finish a job quickly. This is a smart approach, but before that, providing training to employees is necessary otherwise, they may not be able to work in a team together. Too much exposure to heat, or learning of new technologies every day, are factors that cause job burnout in high scale, which gradually leads to employee turnover.

*Problems related to Job Satisfaction*

Oil and gas firms regularly do not provide benefits and incentives to staff. According to Chung et al., (2017), people who work in these organisations risk their lives to finish work. It is, therefore, necessary for all of these organisations to provide health and safety schemes to their employees and managers, otherwise, they may start feeling dissatisfaction in the job. For employees to stay focused on their work, it is essential to provide them benefits like sick leave, paid leave, attendance incentives, etc. High salaries compared to other work areas is not enough; since oil and gas companies generate high revenues from oil exports. The employees must also be offered extra benefits to check and control the employee turnover rate. Moreover, at a few places, employees are forced to work in a field which is not their specialisation, nor are they interested. For example, an employee who has good knowledge of drilling engineering is forcefully asked to work in energy engineering. This causes further job dissatisfaction, which must be curtailed.

*Problems Related to Emotional Intelligence*

In order to stay in any working atmosphere for a longer time by finishing the job in time and by maintaining a good relationship with all, emotional intelligence is required (Lu, Sun, & Du, 2016). However, in reality, there are several organisations, some of which are oil and gas companies, do not recruit employees that know this field. As mentioned above, this is extremely important for organisations to have managers and staff who can show empathy towards others and develop strong bonding with different people. Due to not recruiting these types of people, tasks are not being completed suitably. Employees are losing focus as some staff are not focused on working jointly.

Other oil and gas companies in UAE including Cameron, Petrofac, SONC, Dragon oil, and Technip, etc., are doing their business in similar markets. However, it has been seen that job satisfaction for the employees in the UAE has a higher rate than other countries. Apart from this, these industries are experiencing a high turnover rate in their organisations more than in other countries. To study employees' turnover intention, the independent and dependent variables include job burnout and the emotional intelligence of employees. The mediating variable is job satisfaction, which can mediate the relationship between Job Burnout and Turnover Intention. Imran, Allil, & Mahmoud, (2017) assert that the turnover behaviour of employees is affected by the behaviour of their leaders. Moreover, every employee wants to fulfil their satisfaction level by doing their job in several places. This helps them to meet their everyday needs as well. In this context, job satisfaction levels can mediate the relation between different independent factors and the intention of employees' turnover (Lee, 2018). This study has taken within its scope employees' and managers' opinions of oil and gas companies in the UAE. Approximately 382 workers have been sampled for this study who have participated in the survey process of this research. Moreover, there will be 45 close-ended questions, including those of demographic characteristics.

The selection of IVs and DVs are based on the principle that employees in their workplaces often face the stress that leads to job burnout situation. This stress leads to emotional, mental, and physical exhaustion, resulting in a kind of hopelessness and disappointment, affecting their life and deteriorating their performance levels. Akgunduz & Sanli, (2017) have drawn attention to the fact that stress at the workplace can cause problems related to health as well as emotional outbursts and frequent anger. Chung et al., (2017)emphasise studying the emotional intelligence of employees at their workplaces to examine their actual physical and mental state at workplaces. There are components of emotional intelligence which people need to emphasise, like self-regulation, self-awareness, empathy, motivation, and social skills. Grunfeld et al., (2000) opine that these factors can enhance workers' behaviour in their workplaces and change the style of doing their job.

The mediating variable of this study is the job satisfaction level of employees included in the scope of this study. The study intends to examine whether it can play any mediating role between job burnout and turnover intention. Leunissen et al., (2018) believe that job satisfaction at work helps people to develop their self-motivation, which plays an active role in resolving burnout issues and checking the rate of employee turnover. Hence, this study shall examine the mediating role of job satisfaction with variables like job burnout, emotional intelligence, and turnover intention. The theoretical significance lies in the study of several variables together in relation to employees' turnover intention. These factors help to understand both motivational and stress factors of workplaces. These factors can offer useful insights to leaders and managers of organisations. Lu et al., (2016) state that, owing to some of these factors, employees in the UAE companies prefer to change their jobs. They receive several kinds of benefits from different companies, which helps them to leave their present position to take up a position with more benefits and advantages. There is a deep significance in understanding the relationship and the impact of the IVs and the DVs and the mediating factors of this study. This study shall use a quantitative method of research employing only survey sessions with employees of oil and gas companies in UAE. This will be another theoretical contribution to this subject and will add a new avenue of understanding.

This research is also significant as it is based on different types of data collected through several respondents. It implicates the impact of factors like job burnout, job satisfaction, emotional intelligence, and job turnover intention. The research process itself benefits the researcher, and helps to understand the thinking process of employees in the oil and gas industry and reasons for the employee turnover intentions. Lee, (2018), however, argues that every employee has his or her regular needs and has full freedom to choose a career or a job to work and earn money. The research findings will show how employees focus on better services and better work environment and processes in other companies where they might get more job satisfaction, and hence their decision to turnover. Moreover, this study has used quantitative statistical methods for analysing the findings. These data analysis techniques and the statistical data will serve as terms of reference for future studies. It is also significant to note that such statistical methods increase the validity of data and help researchers to get the beneficial knowledge in their future research as well. About 350 respondents will participate in this research through close-ended questions in relation to job burnout, emotions, and job satisfaction, which can affect their turnover intention.

Additionally, other companies in the UAE market in the gas and oil sector can also benefit from this study. These companies provide a similar work environment to their workers, and the findings of this study will be equally applicable and relevant to their employees as well. These companies can use this information to improve the level of their work environment and overcome their weakness. They can attract more employees to their businesses and increase their employee retention rate. Last but not least, the findings of this study will broaden the understanding of the problem of employee turnover intention since the increase in competition between oil and gas sector companies has also increased issues related to employee retention. This can be seen in the change of mentality of the employee for job turnover too. This study will develop a better understanding of these issues. It is hoped that the companies will start applying some of the strategies and recommendations of this study for retaining their employees. Akgunduz & Sanli (2017) suggest that positive or negative work culture can affect employees' satisfaction levels significantly. If the effect is positive, it enhances their satisfaction level. If it is negative, the satisfaction level decreases and adversely affects their performance levels.

**Literature Review**

Right at the outset, this chapter has discussed several underpinning theories that can be applied by oil and gas companies to overcome their challenges in the retention of employees. Employees of these companies are affected by extreme pressure from their workplaces, this results in several health-related problems, and many of them get frustrated at work. This phenomenon is considered as job burnout, which can affect the productivity of workers and slow down their working performances and working speed. This issue can be partially resolved by the maintenance of the emotional intelligence of employees to control the turnover intention of employees.

**Job Burnout**

Job Burnout is defined as a state where an employee of an organisation feels mental, physical, or emotional stress, and which gradually leads to depression. Wright (2011) observes that stress is caused when organisations pressure employees to perform extra work or to perform actions outside of their fields. In all theseinsttances, employees start feeling bored and overworked, which makes them anxious in the working environment. Moreover, as argued by Purvis, Zagenczyk, & McCray (2015), when an employee performs the same tasks each day, then that person loses motivation to come to work as there is no challenge in completing tasks. All these reasons cause demotivation and develop negative energy inside employees, and job burnout takes place.

Job burnout is becoming a common factor in several different careers and causes employee turnover to increase. According to a recent study, nearly 23% of employees burn out very often or almost every day at work, while another 44% of employees feel burned out at times. This implies that very few employees do not experience burnout. However, organisations are not taking any step to stop that. Gharakhani & Zaferanchi (2019) found out that due to job burnout, several health issues take place among employees including coronary heart diseases, high cholesterol, and gastrointestinal issues. Several employees die due to job burnout, and they are aged below 45 years.

There are several reasons behind job burnout including a lack of support from higher authority, unfair deadlines, extreme workload, and pressure from the boss. Moreover, several organisations even give employees work after they return home from the office, and this consumes the personal lives of those employees. Steffens, Yang, Jetten, Haslam, & Lipponen (2018) mentions a few other causes of job burnout like a lack of promotion advantages, monotony, and an inability to see results. Employees often feel burnout when they are stagnant in a position for a long time or when they do not get promotions. This is the current situation of Job Burnout in the UAE.

Several past researchers have examined job burnout situations among employees of different oil and gas companies in the United Arab Emirates. According to Allen, Peltokorpi, & Rubenstein (2016), the primary reason behind job burnout in the UAE is less availability of different career enhancement opportunities. Nearly 64% of employees are not satisfied with the job opportunities provided to them. Employees, even after working hard, do not get any chance to move ahead with the implementation of a successful career or professional lives. For that reason, they start feeling frustrated, and slowly they start feeling stressed. No promotion for a long time can make them doubt their abilities and whether they are appreciated.

Oil and gas companies are required to be more open about giving promotions to eligible employees as they work with a considerable amount of risks every day. Moreover, rude colleagues, managers, and long hours catalyse feelings of extreme stressed for employees who do not get any promotion or increment, resulting in employee turnover. On the other hand, as contradicted by Osabiya (2015), organisations pressure employees on a large scale regularly. Pressure is not anything harmful and if it is used at the right time, then employees can get motivated to work. However, different organisations, especially oil and gas companies in the UAE, do not know how to use pressure as a motivational tool. 45% of employees of the UAE are satisfied with the amount of pressure they get from higher authority, while 16% only are delighted. Thus, as stated by Yunus & Kamal (2016), organisations must understand which employee can take of the pressure, and then they are required to allocate tasks based on that observation. However, on the other hand, as contradicted by Neubauer & Martskvishvili (2018), managers can develop or destroy employees.

Out of all employees who work in oil and gas companies in the UAE, even less than 50% have stated that they are happy with their managers. Only 19% of people have indicated that they get massive support from their managers. As contradicted by Najjar & Fares (2017), organisations are required to understand the capabilities, skills, and withstand the power of employees. Moreover, they are also required to realise when employees are feeling job burnout so that they can give a break to them.

There are several different symptoms of job burnout like sudden outbursts due to temper, depression, panic, and demotivation. Moreover, as opposed by Yahya et al. (2018), there are also burnout components or stages like emotional exhaustion, depersonalisation, and decrease in accomplishment. All these stages are required to be understood by employers so that they can help employees to reduce job burnout.

**Emotional Intelligence**

Emotional Intelligence is referred to as the ability of any person to control their own emotions as well as control other emotions. According to Mayer, Caruso, & Salovey (2016), employees and managers of any organisation must have the ability to control and influence the emotions of others. Rather than employees, especially for managers and leaders, it is necessary to have this skill. Leaders are ones who control and manage a team. They are responsible for whatever happens inside that team and with their team members. As stated by Andrei, Siegling, Aloe, Baldaro, & Petrides (2016), employees or leaders are both required to have emotional intelligence, so that they can give support to other team members and work together as a great team.

In general, five components of emotional intelligence must be inside every leader and employee sometimes as well. According to Cho, Rutherford, Friend, Hamwi, & Park (2017), the essential elements of emotional intelligence are self-awareness, motivation, self-regulation, social skills, and empathy. Self-awareness helps people to understand situations by which they are going through so that they can guide others who are going through the same situations. Self-awareness, in short, means that leaders or employees are aware of their weaknesses and strengths. On the other hand, as argued by Lloyd, Boer, & Voelpel (2017), motivated people can perform tasks more quickly and efficiently than anyone else. They have ideas about what is required to be done, and they do it by adequately guiding their teams.

Moreover, another important factor here is known as self-regulation. As opposed by Mayer et al. (2016), self-regulation helps people to think before they speak, and it is an essential aspect to consider. Different leaders, managers, and even employees must have ideas about self-regulation so that they can strategically complete work. Another factor is empathy, and it is useful as employees and leaders must show compassion towards each other. Social skills deal with communicating points of view of any person to others so that everyone can have ideas about what each member of any organisation thinks.

Several past types of research have been done based on the impact of emotional intelligence on organisations and employees of oil and gas companies in the UAE. According to Cho et al. (2017), to stay longer in any organisation, emotional intelligence is essential for both employees and leaders or managers. Out of all factors of emotional intelligence, the most important and interesting one is self-awareness. This particular one deals with making people understand and realise their positions or the situations through which they are passing. In the case of the UAE, developing and exporting oil is the most critical work that is done by this country.

Thus, a considerable amount of organisations are related to this job, and different employees and leaders join this industry every year. They are required to have ideas about emotional intelligence to work with thousands of employees inside an organisation. Self-awareness makes them understand the current situation and let them develop solutions to mitigate issues inside the working atmosphere. On the other hand, as argued by Li (2015), self-regulation helps people to understand the value of thinking before speaking. Several times managers of oil and gas companies in the UAE speak before thinking twice, and for that reason, employee turnover is increasing as employees are not satisfied with higher authorities. Self-regulation will help managers to speak well without hurting anyone.

Moreover, as argued by Andrei et al. (2016), employees will also be able to admit their own mistakes and stay calm by self-regulation. Different oil and gas companies of the UAE set goals and motivate employees to achieve that. However, sometimes motivation can be at stake due to job dissatisfaction and job burnout of employees. In all these cases, managers or leaders of the organisation are required to show empathy towards employees and understand their situations to make them feel better. On the other hand, as contradicted by Pokropski (2017), managers must have social skills as by that they can communicate what they need to employees. Moreover, employees will also be able to communicate with others by having these skills. Thus, in the case of the oil and gas companies of the UAE, every employee and manager is required to have emotional intelligence.

**Job Satisfaction**

Job satisfaction can be considered as an extent by which employees can experience self-motivated, satisfaction, and happiness with their workplaces and work process. In accordance with Alfayad & Arif (2017), different factors of companies are related to providing satisfaction to workers by which they can increase their business performances as well. Furthermore, within workplaces, stability and feeling comfortable are other aspects which help employees feel satisfied. In this case, job satisfaction can be stated as one of the components of organisational behaviour, and this enhances the fulfillment of the needs of any person related to organisations.

In the case of the oil and gas industry of UAE, there are different companies doing their business in this sector, and these companies provide most of the economic factors to this nation. As stated by Yunus & Kamal (2016), the satisfaction of a job is a complex structure, and this is related to feelings in a personal manner and an effective impact on the working process of employees. Moreover, job satisfaction can be linked up with the intention of turnover between employees of business organisations. On the other hand, organisational job satisfaction is a related factor in their job-related perspective. This has seen companies of the oil and gas sector face several changes in the prices of their products as per the economic aspects of nations. In this context, most of the companies also require to change their financial factors. In accordance with Steffens et al. (2018), this effect in job satisfaction level of employees who enhance their intention to leave their job for getting more benefits from other companies and workplaces.

Business companies can put several efforts into their business to increase the productivity of their workers in a short time. Moreover, the HRM team of companies is required to focus on this factor in a more relevant manner, which helps them to retain their workers in a specific way. In accordance with Schleicher, Smith, Casper, Watt, & Greguras (2015), this is related to organisational culture also, and people can feel comfortable with a positive organisational environment, which helps them to increase their satisfaction level in their workplaces.

This has seen that job satisfaction can be different in male and female employees of the oil and gas industry. Moreover, the satisfaction level can be based on age, sex, culture, and other demographic levels. In accordance with Allen et al. (2016), several employees of companies are satisfied with average beneficial factors. Moreover, these people can reliably do their work and focus on their work process to develop their work performances. On the other hand, many people and employees require high satisfaction from their companies. In this case, the organisations of the oil and gas sectors need to provide high-quality benefits to their employees for training them. This decreases the turnover intention of employees as they appropriately meet their requirements.

In a study of the oil and gas industry, the researchers have seen that satisfaction level were slightly higher for the male workers of oil and gas sectors. In this case, 26.3% of male employees are satisfied with their work. On the contrary, only 24.4% of female employees of this similar industry have similar satisfaction levels to their male employees in relation to the same benefits from their organisation (Casper & Buffardi, 2004). In relation to the business market of the UAE, most of the employees become satisfied with their workplaces and their benefits. Among several countries, the level of job satisfaction in employees of UAE is highest, which has known by surveying 1000 employees in different 128 countries.

This has seen for the UAE market that there are only 31% of people who are seeking full-time employment in any business. Furthermore, 58% of people who are already employed in a full-time manner rated their workplaces and work procedure as the code of 'good,' and the other 1% have stated as 'great.' These percentages are higher than the percentages of other countries such as Russia and the U.S (Tanova & Holtom, 2008). In relation to financial factors, most of the places in this region increase their salaries in a relevant way. This has seen that companies have increased their salary in a specific percentage to retain their workers. In this manner, companies of the UAE have increased their salary by approximately 17% and this help employees to fulfil their requirements positively. In relation to the satisfaction level of employees, one can apply Herzberg's theory in their work process which helps them to differentiate between hygiene and motivational factors. Motivation from the leaders and managers can enhance comfort ability of people in between employers and employees relationship. As argued by Osabiya (2015), employees need to motivate themselves as. Maslow's motivational theory can state this in better manner. However, in the case of oil and gas sector of the UAE, this industry of Abu Dhabi has approximately 94% of researchers of oil than other states. This industry is relatable with the UAE economy and provides several benefits to its employees for their high demand. Moreover, many employees have an intention to change their workplaces in similar sectors also. There are several other factors of business which can influence and affect the satisfaction level of workers to increase their productivity and performances. On the other hand, as influence from this can be stated that job satisfaction from any company can mediate intention of turnover of employees (Huang et al., 2016).

**Turnover Intention**

Turnover intention is related to employees of organisations in which they plan to leave their current job for changing their workplace. Moreover, this intention of employees is dependent upon several factors of business companies. In accordance to (Gharakhani & Zaferanchi, 2019), job burnout, job satisfaction level, psychological contract between employers and employee can affect the turnover intention of employees in both positive and negative manner. Furthermore, commitment from organisation, appreciative leadership and work environment of companies can affect the behavior of employees in relation to leaving their job as well. On the other hand, as argued by (Snipes, Oswald, LaTour, & Armenakis, 2005), job satisfaction level which employees get from their business company can affect and mediate the other factors in both positive and negative way and this also changes the effect for job turnover intention. However, turnover intention can be related to business companies as well in which many organisations plan to terminate their employees from their workplace position.

In this context, employee turnover intention can be stated as dependent variable which is based on the different independent variables in relation to both employee and organisation. Moreover, as opined by Holman & Axtell (2016), turnover intention of workers become high when they get opportunity to get more benefits from other companies in similar sector with similar effort. This has seen that high turnover rate of business companies can affect their performance level in relevant manner. This can decrease popularity of companies which affects their business negatively. Oil and gas companies of the UAE are doing their work in risky situations. In this case, these kinds of companies need to provide different benefits to their employees for retaining them. This has seen that the UAE has ranked 2 in its economical aspect after Saudi Arabia. The petrochemical industry is known as most important industry for their nation for effect in their GDP increment process.

This nation has known to hold more than 97.8 billion barrels of oil in its reserve. In this context, 96% reserve in present in Abu Dhabi and Dubai. Apart from this, the natural gas sector of the UAE can consider as another leading sector for UAE economy. This has seen that the gas sector of the UAE market uses approximately 215 trillion cubic feet place in overall UAE for maintaining their business. There is more than 50 companies dealing in the oil and gas sector of the UAE. This provides a high amount of revenue to this nation for which they need a huge number of labourers for managing their business. In this case, as opined by Oerlemans & Bakker (2018), competition can be faced by companies which are doing their business in similar industry. In the context of managing demand of the oil and gas sector, these companies provide relevant benefits to their workers for retaining them. In 2015, the UAE has produced approximately 3179 BCF of the gas in their natural gas sector. In the case of employee turnover, UAE has increased its employee turnover rate in different business approximately 23.4% in 2018. Furthermore, this is increased from 20.6% in 2012. On the other hand, in 2015 there are 31% of employees have changed their companies and workplaces in 2015. Additionally, there were 57% of people who have turnover intention from their workplaces in 2016. This is showing higher intention rate for turnover in UAE market.

The hypotheses for this research paper are as follows:

- **Hypothesis 1:** The job burnout construct has a positive relationship to turnover intention
- **Hypothesis 2:** The job satisfaction construct has a positive relationship to turnover intention
- **Hypothesis 3:** The emotional intelligence construct has a positive relationship to turnover intention
- **Hypothesis 4:** Job Satisfaction mediates the relationship between Job Burnout and Turnover Intention

**Research Methods**

The present study adopted a quantitative data collection and analysis process. In this regard, around 386 respondents had been chosen from oil and gas companies in the UAE. The total employees that are currently working in oil and gas companies in the UAE total 55,000. In this regard, based on Krejcie and Morgan's (DATE) table, there is a minimum of 386 respondents selected for the analysis process of this study. A simple random sampling technique has been adopted by the survey for choosing 386 samples in the survey session. After getting the formal permission, 386 questionnaires were distributed to the employees of oil and gas companies in the UAE, but out of them, only 293 were returned and used for the analysis process and had 73.98 percent response rate. The job burnout (JB) has five items, emotions (EM) has five items, job satisfaction (JS) has five items, and turnover intentions (TI) also has five items. These are shown in Figure 1.

**Figure 1:** Theoretical Framework

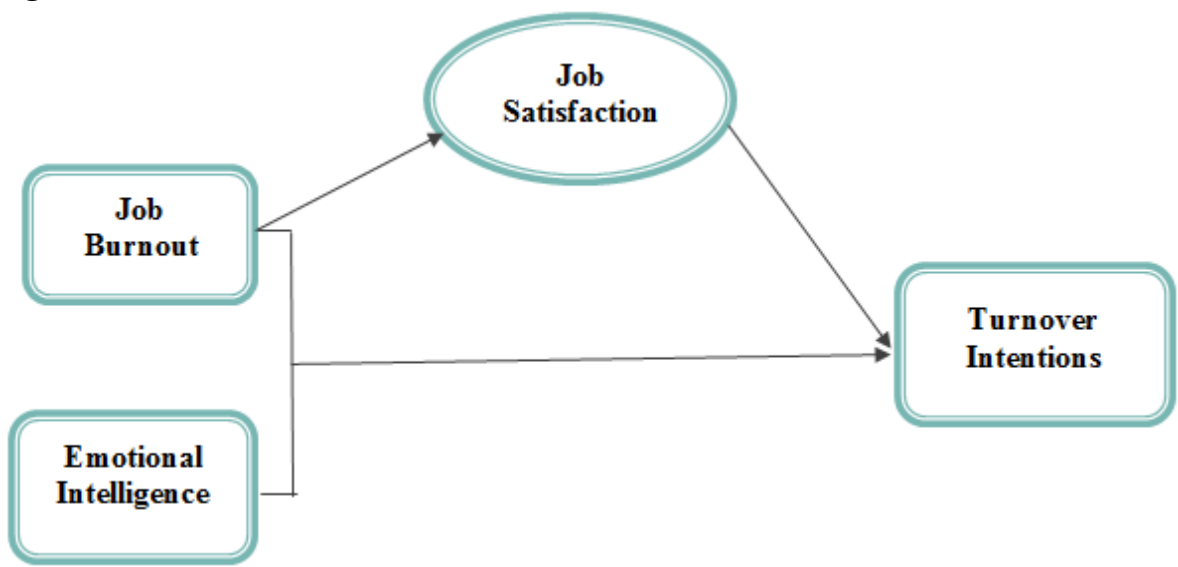

**Findings**

The results of the ongoing study firstly shows the convergent validity that describes the links among the items and the statistics of loading and AVE are larger than 0.50 while statistics of CR and Alpha are not lower than 0.70 which shows that there is a high linkage among the items that are highlighted in Table 1 given below:

**Table 1:** Convergent Validity

| Constructs | Items | Loadings | Alpha | CR | AVE |
|---|---|---|---|---|---|
| Emotions | EM1 | 0.807 | 0.840 | 0.893 | 0.676 |
|  | EM2 | 0.846 |  |  |  |
|  | EM4 | 0.826 |  |  |  |
|  | EM5 | 0.808 |  |  |  |
| Job Burnout | JB1 | 0.800 | 0.872 | 0.908 | 0.663 |
|  | JB2 | 0.878 |  |  |  |
|  | JB3 | 0.855 |  |  |  |
|  | JB4 | 0.757 |  |  |  |
|  | JB5 | 0.775 |  |  |  |
| Job Satisfaction | JS1 | 0.816 | 0.861 | 0.900 | 0.642 |
|  | JS2 | 0.861 |  |  |  |
|  | JS3 | 0.744 |  |  |  |
|  | JS4 | 0.743 |  |  |  |
|  | JS5 | 0.836 |  |  |  |
| Turnover Intentions | TI1 | 0.796 | 0.849 | 0.898 | 0.689 |
|  | TI2 | 0.836 |  |  |  |
|  | TI4 | 0.850 |  |  |  |
|  | TI5 | 0.835 |  |  |  |

The output of the current study secondly show the discriminant validity that describes the nexus among the constructs and the figures of Heterotrait Monotrait ratios are not bigger than 0.90 that show the no high linkage among the constructs that are highlighted in Table 2 given below:

**Table 2:** Heterotrait Monotrait Ratio

|    | EM    | JB    | JS    | TI |
|----|-------|-------|-------|----|
| EM |       |       |       |    |
| JB | 0.587 |       |       |    |
| JS | 0.630 | 0.493 |       |    |
| TI | 0.737 | 0.664 | 0.631 |    |

**Figure 2:** Measurement Model Assessment

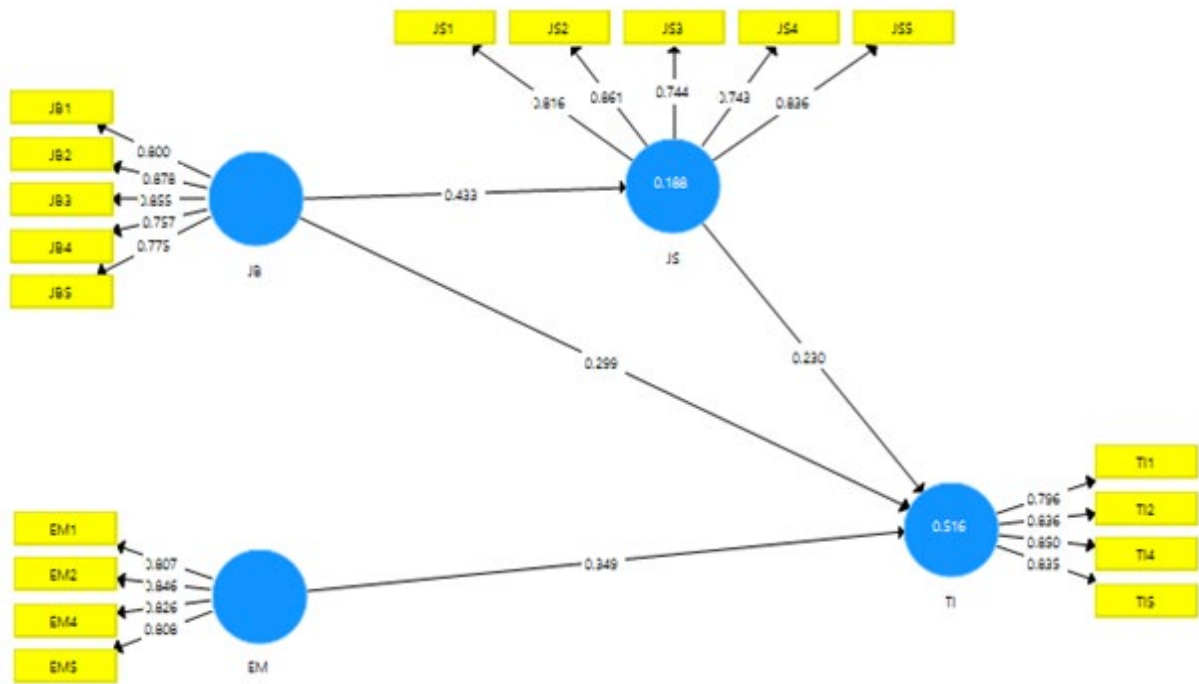

The findings of the path analysis show that job burnout, job satisfaction, and emotional intelligence have a positive association with turnover intentions and accept H1, H2, and H3. In addition, job satisfaction has positive mediation among the links of job burnout and turnover intentions and accept H4. These path analyses are highlighted in Table 3.

**Table 3:** Path Analysis

|                | Beta  | S.D.  | t-values | p-values |
|----------------|-------|-------|----------|----------|
| EM -> TI       | 0.349 | 0.034 | 10.365   | 0.000    |
| JB -> JS       | 0.433 | 0.034 | 12.793   | 0.000    |
| JB -> TI       | 0.299 | 0.035 | 8.598    | 0.000    |
| JS -> TI       | 0.230 | 0.033 | 6.873    | 0.000    |
| JB -> JS -> TI | 0.100 | 0.016 | 6.241    | 0.000    |

**Figure 3:** Structural Model Assessment

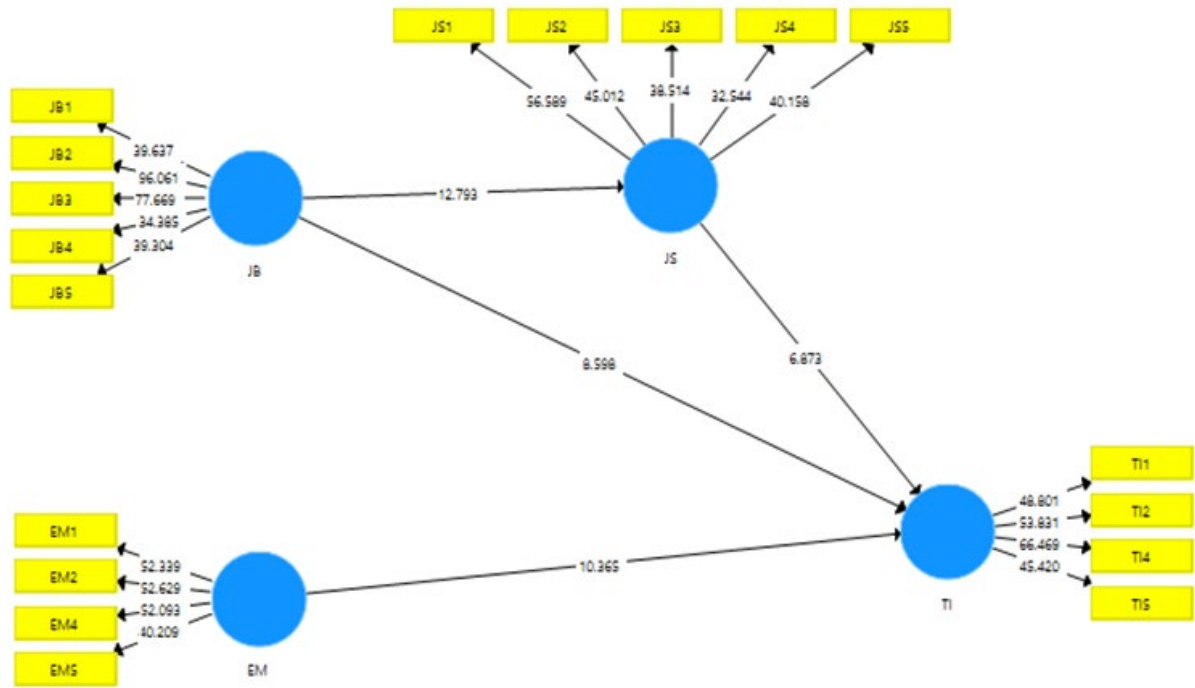

**Discussions and Conclusion**

This study has examined the relationship between job burnout, job satisfaction, emotional intelligence, and turnover intention in oil and gas companies in the UAE. This study aims to investigate the mediating effect of job satisfaction on the relationship between job burnout and turnover intention. In addition to this, the present research has empirically tested the mediating role of job satisfaction between job burnout and turnover intention, considering the same sample of the study. To address this relationship, the present study was based on the four research questions and research objectives. The research objectives were developed to test the direct link between job burnout and turnover intention (RO1), between job satisfaction and turnover intention (RO2), emotional intelligence and turnover intention (RO3) while to test the indirect relationship of job satisfaction among the links of job burnout and turnover intention (RO4). To address all of these research objectives, overall, four research hypotheses were developed and tested through Smart-PLS. The research hypotheses under the title of H1-H3 have tested the direct relationship between the variables. It was found that a significant relationship exists between job burnout and turnover intention, between job satisfaction and turnover intention, between emotional intelligence and turnover intention. All these relationships through empirical evidence are proved. The research hypotheses for the direct relationship were accepted. After testing the direct relationship between the variables, R4 was developed to test the mediating effect of job satisfaction and the above-stated variables. The findings through the mediating effect indicate that H4 was justified with the empirical findings. More specifically, it is found that job satisfaction positively mediates among the relationship between job burnout and turnover intention. These results provide the guidelines to the policymakers that they should enhance their focus on the job satisfaction and turnover intentions of the employees that improve the firm performance.

**Conclusion**

Finally, this study concluded job satisfaction is playing a vital role in the intentions of turnover among the employees in the oil and gas companies. This study is useful evidence for the firms working in the Oil and Gas sector in the UAE to overcome the issue of high turnover intention due to the direct effect from exogenous variables. In addition, the mediating role of job satisfaction may also be viewed as an excellent contribution to the literature. In terms of higher business practices and strong governance structure, this study might be utilised as a practical guideline to improve the employee's retention while addressing their core issues and problems. With effective management expertise, managers, and other decision-makers in the selected companies can reduce the environment of uncertainties and uneven future results with more satisfaction from their employees. Through identifying the significance of exogenous variables in defining the endogenous variable (turnover intention), the overall corporate environment will move in a positive direction. Furthermore, this approach could help to better some strong business strategies and for employees too.

**Limitations and Future Directions**

The first limitation indicates the coverage of only Oil and Gas Companies in the UAE. This fact would show that the present study is limited in terms of one industry, which seems to be insufficient for the generalisation of the results. Additionally, this study also has covered a sample of 290 respondents, which is found to be reasonable but not very good enough to generalise the study output. It is better to expand the sample size in future studies, which may provide some good results, compared to current statistical results. Meanwhile, the study has observed the mediating effect of job satisfaction between job burnout and turnover intention of the employees. Future studies are highly recommended to expand the mediating effect of job satisfaction between emotional intelligence, commitment, psychological contract, and quality of the work-life with turnover intention.